\documentclass[final,1p,times]{elsarticle} 
\usepackage{graphicx}
\usepackage{amssymb} 
\usepackage{amsthm} 
\usepackage{lineno}

\journal{Nuclear Physics A} 

\begin{document}

\begin{frontmatter} 

% Your Title - please insert
\title{Quantum chaos in the perfect fluid:\\ spectrum of initial fluctuations in the little bang}

%% Single author (and collaboration) - please insert
\author[auth1]{Kevin Dusling}
\author[auth2]{Thomas Epelbaum}
\author[auth2]{Fran\c cois Gelis}
\author[auth3]{Raju Venugopalan}

\address[auth1]{Department of Physics, North Carolina State University,  Raleigh, NC 27695, USA}
\address[auth2]{Institut de Physique Th\'eorique (URA 2306 du CNRS)
  CEA/DSM/Saclay 91191, Gif-sur-Yvette Cedex, France}
\address[auth3]{Physics Department, Brookhaven National Laboratory, Upton, NY 11973, USA}

\begin{abstract} 
We outline how unstable quantum fluctuations decohere classical fields in heavy ion collisions, leading to an equation of state and hydrodynamics. Explicit numerical realization of this framework in a scalar $\phi^4$ theory demonstrates that anomalously low values of $\eta/s$ can be generated. 

\end{abstract} 

\end{frontmatter} % do not change

%% linenumbers are useful for reviewing process
\linenumbers

The motivation for this work is to achieve a deeper understanding of what one might consider the unreasonable effectiveness of hydrodynamics in heavy ion collisions. Hydrodynamics is a good effective field theory for the late time long wavelength behavior of a quantum field theory~\cite{Arnold:1997gh}. Why does it appear to work so well at times $\leq$ 1 fm in heavy ion collisions ? Another important motivation is to compute the right non-equilibrium initial conditions that can be matched on to viscous hydrodynamics, thereby eliminating an {\it ad hoc} feature of current phenomenology. 

Our approach is a weak coupling one, albeit the dynamics is very non-perturbative. Abundant analogies to such dynamics exist in other fields of physics. In this Color Glass Condensate effective field theory~\cite{CGC}, quantum fluctuations can be isolated and computed in principle order by order in $\alpha_S$; each order in this expansion includes resummations of different contributions depending on the nature of these fluctuations. Before the collision, we have to worry about factorization of quantum fluctuations into the wavefunctions of the incoming nuclei. The proper treatment of these is important to understand the energy evolution of the wavefunctions. Factorization implies that the rapidity $Y$-dependence of the density matrix $W_Y[\rho]$ describing the n-body correlations of color charge densities $\rho$ in the nuclear wavefunctions--described by the JIMWLK Hamiltonian~\cite{JIMWLK}--is universal, regardless of whether the high energy probe of the nucleus is an electron, a proton or another nucleus. If true, this universality  (proven currently~\cite{Factorization} only for ``leading logs'' in $\alpha_S Y$ at each order of perturbation theory) would be a powerful predictor of a wide range of phenomena in high energy QCD. 

The focus here is on heavy ion collisions, where the leading order description is in terms of collisions of classical fields which best describe the high occupancy gluon fields in the nuclear wavefunctions~\cite{MV}. In QCD one has classical gluon production, which gives the dominant contribution at early times~\cite{CYM}. At early times, this matter, called the Glasma~\cite{Glasma}, contains lumpy (of size $1/Q_S$, where $Q_S$ is the saturation scale) configurations of strong longitudinal chomo-electric and chromo-magnetic fields, giving rise to  very anisotropic configurations with pressures 
$P_T >>P_L\sim 0$~\cite{Yang-Mills}. In gauge theories, these are known to give rise to instabilities, either of the Weibel~\cite{Weibel}  or Nielsen-Olesen type~\cite{Nielsen-Olesen}. Therefore small $O(1)$ quantum fluctuations can grow to be as large as the background classical fields~\cite{Romatschke:2005pm} on parametric time scales (for expanding systems) $\tau\sim \ln^2(1/\alpha_S)/Q_S$. 
Since such contributions can occur at each order of perturbation theory, a resummation of these is required to achieve stable results. 

We showed recently that leading temporal instabilities (those that grow as ($\alpha_S\exp(2\sqrt{Q_S\tau}))^n$, where $n$ is an integer denoting the order in perturbation theory) can be resummed and expressed in terms of a gauge invariant spectrum of fluctuations on an initial Cauchy surface at $\tau=0^+$~\cite{Dusling:2011rz}. This work suggests that the temporal evolution of inclusive quantities (such as the stress-energy tensor $T^{\mu\nu}$ or correlators thereoff) can be expressed in terms of a ``Master 
formula" 
\begin{equation}
\langle T^{\mu\nu}\rangle_{{\rm LLx + LInst.}} =\int [D\rho_1 D\rho_2]\;
W_{x_1}[\rho_1]\, W_{x_2}[\rho_2]  \int \!\! \big[{\cal D}\alpha\big]\,
F_0\big[\alpha\big]\; T_{_{\rm LO}}^{\mu\nu} [{\cal A}[\rho_1,\rho_2] + \alpha] (x)\; , 
\label{eq:final-formula}
\end{equation} 	
where the argument ${\cal A}\equiv ( A, E)$ denotes collectively the components of the classical fields and their canonically conjugate momenta on the initial proper time surface; analytical expressions for these are available at
$\tau=0^+$~\cite{CYM}. Their temporal evolution is obtained by solving  Yang-Mills equations~\cite{Yang-Mills}. The $W$'s are the functional density matrices defined previously that obey the JIMWLK equation.

The initial spectrum of fluctuations $F_0\big[\alpha\big]$, Gaussian in the quantum fluctuations $\alpha$,  has a variance given by the small fluctuation propagator in the Glasma background field at $\tau\rightarrow 0^+$. In practice, the path integral in $\alpha$ is determined by solving the classical Yang-Mills equations repeatedly with the initial conditions at $\tau=0^+$ given by
\begin{equation}
{\bf A}_{\rm init.}^\mu = {\cal A}_{\rm init.}^\mu + \int d\mu_{_K}\;\Big[c_{_K}\,a_{_K}^\mu(x)+c_{_K}^*\,a_{_K}^{\mu*}(x)\Big] \, .
\label{eq:quantum}
\end{equation}
Here ${\bf A}^\mu$ denotes the quantum fields and their canonical conjugate momenta. The  coefficients $c_{_K}$, with $K$ denoting the quantum numbers labeling the basis of solutions, are random Gaussian-distributed complex numbers. Explicit expressions for the small fluctuations and their conjugate momenta, denoted here by $a_K^\mu (x)$ were obtained in ~\cite{Dusling:2011rz}. These need to be evaluated numerically--a challenging enterprise--but significant progress has been made in this direction. 

Considerable insight is gained by formulating the analogous problem in a massless scalar $\phi^4$ theory~\cite{fixed1}. This theory, like classical Yang-Mills, is conformal and has instabilities due to parametric resonance of quantum fluctuations with the classical background. In the simple example of a fixed box, one sees that the leading order classical energy density and pressure for this theory are not single valued, with the pressure fluctuating rapidly as function of time. However, after adding fluctuations with Gaussian random coefficients (as in Eq.~(\ref{eq:quantum})), the energy density and pressure rapidly develop a single valued relationship, namely, an equation of state (EOS). 

The EOS develops as a result of decoherence. Each of the trajectories corresponding to an initialization of the scalar analog of Eq.~(\ref{eq:quantum}) has a slightly different amplitude. In a conformal theory, different amplitudes have slighly different phases; for spatially independent  fields and fluctuations in the $\phi^4$ theory, one can show that $T_{\rm period} = 18.2/g\Delta \phi_{\rm max}$. In this case, the phase space density rapidly fills the constant energy Poincare surface uniformly; a simple exercise shows that this makes the stress-energy tensor traceless. (The canonical definition of the stress-energy tensor for the scalar theory is not traceless.) For an expanding 1+1-D scalar theory, one obtains Bjorken hydrodynamics. 

It is also instructive to look at spectral functions~\cite{fixed2} obtained from a Fourier transform of the imaginary part of the retarded Green function\footnote{The resummed expression for the latter is obtained from an equation analogous to Eq.~(\ref{eq:final-formula}). The leading order expression is the fluctuation field that obeys the equation $\left[\square_x + V^{\prime\prime}(\phi(x))\right]a(x)=0$.}. At early times, no quasi-particle behavior is seen, but it develops and one can extract a plasmon mass from the spectral function. Likewise, one obtains the occupation number from the resummed symmetric Green function (the sum of the Wightman functions $G_{-+} + G_{+-}$). This develops a thermal structure $f_k = T/(\omega_k-\mu) - 1/2$, with the $-1/2$ denoting the contribution from vacuum fluctuations. Notably, there is a zero mode that is overoccupied relative to the thermal spectrum: it has been checked~\cite{fixed2} that it  demonstrates the characteristics of a Bose-Einstein Condensate, as also argued elsewhere~\cite{Berges:2012us}. It is speculated that a transient condensate forms in the 
gauge theory analog as well~\cite{BEC}.  

\begin{figure}[tb]
  \begin{center}
    \begin{minipage}{0.495\textwidth}
    \includegraphics[width=7cm]{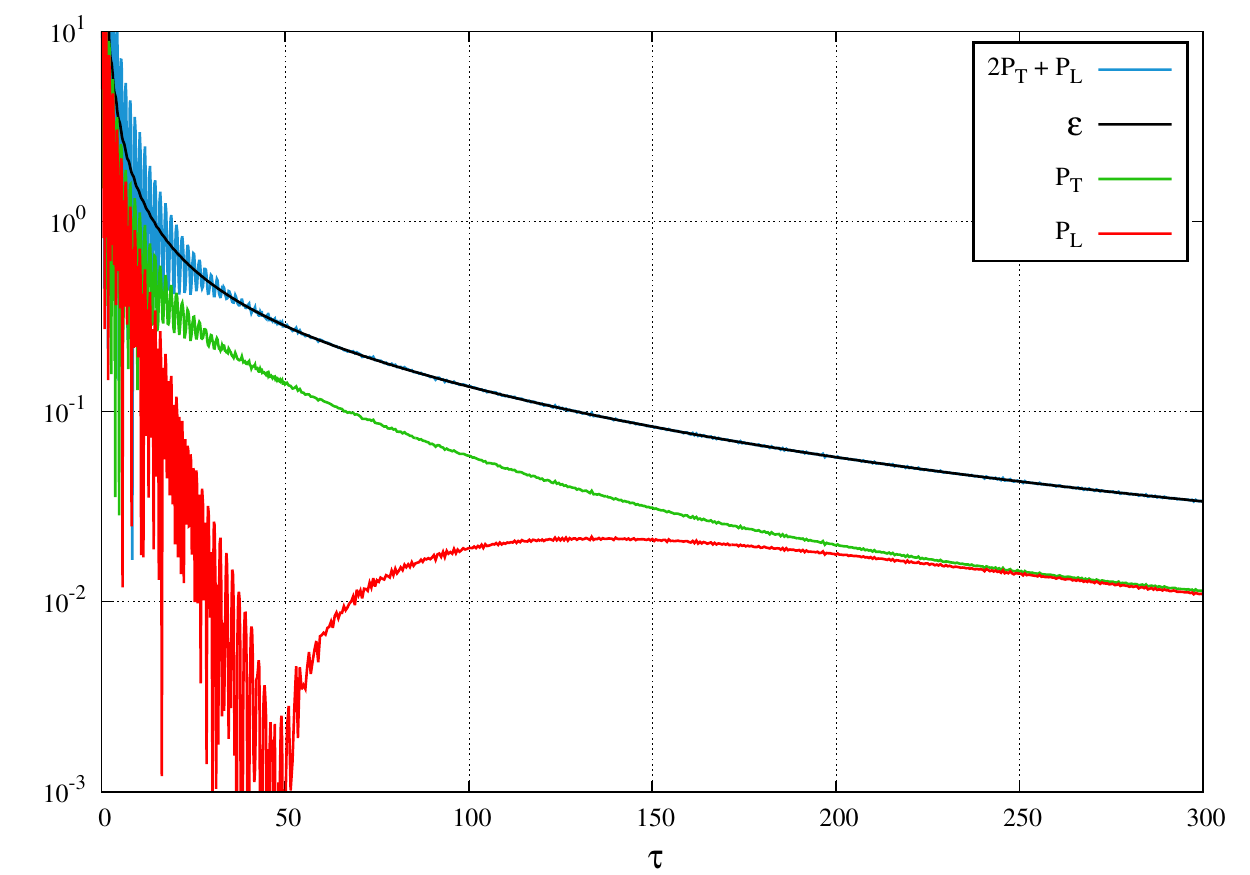} 
    \end{minipage}
    \hfill
    \begin{minipage}{0.495\textwidth}
    \includegraphics[width=7cm]{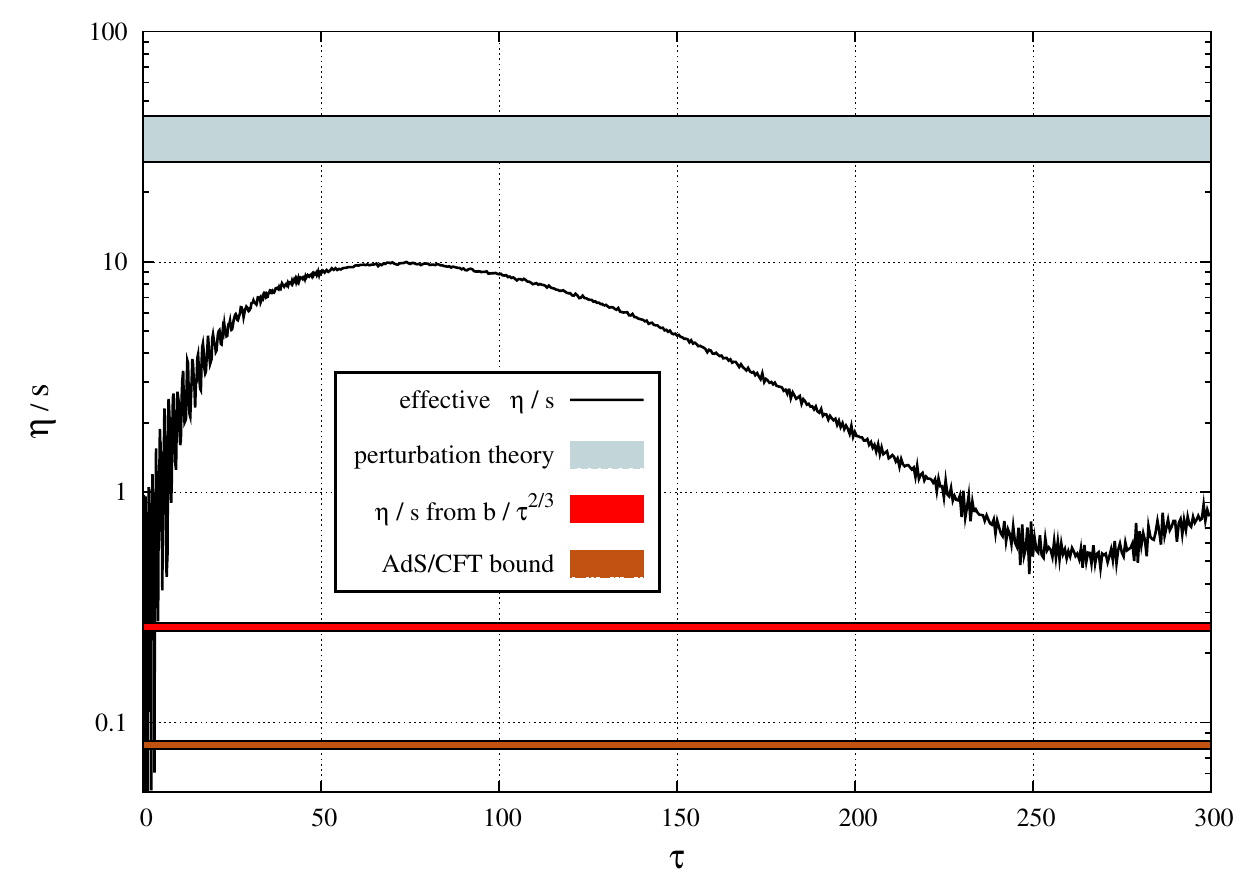}
    \end{minipage}
    \caption{Left figure: Time evolution of absolute values of diagonal elements the stress-energy tensor and the trace of the pressure tensor.  Right figure: Evolution of the numerically extracted $\eta/s$
  ratio, compared to the perturbative value of this ratio, the value extracted from matching to first order viscous hydrodynamics and the conjectured universal lower bound $\eta/s=1/4\pi$.
      \label{fig:one}}
  \end{center}
\end{figure}

The $\phi^4$ analogy to heavy ion collisions--of longitudinally expanding scalar fields--was explored in ~\cite{DusliEGV2}. The theoretical framework is identical. One observes the pattern: Decoherence $\longrightarrow$ EOS $\longrightarrow$ Isotropization. Two striking features are  illustrated in Fig.~\ref{fig:one}. Firstly, one notices (left figure) that despite a rapid red shift of the longitudinal pressure, the explosive instability growth allows the system to beat the expansion, leading to near isotropization at late times. It is important to note that the x-axis is in arbitrary lattice units\footnote{Because of a logarithmic ultraviolet divergence in dynamically generated  $m^2 \phi^2$ terms in components of $T^{\mu\nu}$, composite operator renormalization is needed to relate computations at different lattice spacing and give physical meaning to time scales. Gauge invariance ensures this problem does not exist for QCD.}. 

The other striking result is shown in Fig.~\ref{fig:one} (right), where an effective extracted time dependent $\eta/s$ is shown. This is obtained by matching values of the diagonal components of the stress-energy tensor to first order viscous hydrodynamics using 
\begin{eqnarray}
P_{_T}= \frac{\epsilon}{3}+\frac{2\eta}{3\tau}\;\; ; \;\;
P_{_L}= \frac{\epsilon}{3}-\frac{4\eta}{3\tau}\; .
\label{eq:visc-P}
\end{eqnarray}
To extract $\eta/s$, we use the Stefan-Boltzmann formula to estimate the energy and entropy density, $\epsilon = \frac{\pi^2 T^4}{30}$ and $s = \frac{2\pi^2 T^3}{45}$ with $s \approx \epsilon^{3/4}$. 
In the hydrodynamical regime, 
\begin{equation}
\left[\frac{P_{_T}-P_{_L}}{\epsilon}\right]_{\rm hydro}
=2\,\frac{\eta}{s}\,\frac{s}{\tau\epsilon}\approx 
\underbrace{\frac{\eta}{s}\,\frac{2}{A^{1/4}}}_{b}\;\frac{1}{\tau^{2/3}}\; ,
\end{equation}
where $A$ is the coefficient in the asymptotic behavior of the
energy density, $\epsilon\approx A \tau^{-4/3}$. From this formula, we extract $\eta/s\approx 0.26$. There is a large systematic uncertainty because $(P_{_T}-P_{_L})/\epsilon$ at early times is best fit by a form $\exp(-{\rm Const.}\tau^2)$--a faster relaxation of the pressure anisotropy than 
achieved through the power law behavior $\propto 1/\tau^{2/3}$ typical in hydrodynamics. Nevertheless, the value extracted is {\it two orders of magnitude lower} 
the perturbative value~\cite{Jeon2} $\eta/s\sim 10^4/g^4\sim 40$ for $g=4$ and about three times the conjectured AdS/CFT bound. Such anomalously low viscosities also occur in turbulent phenomena~\cite{AsakaBM} where momentum transport occurs as if the viscosity were much smaller than transport cross-section estimates.

It may be interesting to relate these results to work in the AdS/CFT framework describing potentially universal features in the relaxation of off-equilibrium strongly coupled systems to hydrodynamics~\cite{HelleJW}. Further, numerical simulations underway in the gauge theory case will improve state-of-the art Glasma+hydro computations~\cite{STV} by matching of the dynamics of the Glasma to viscous hydrodynamics.

\section*{Acknowledgments}
 K.D and R.V are respectivey supported under DOE Contract Nos.DE-FG02-03ER41260 and DE-AC02-98CH10886. F.G and T.E are supported by Agence Nationale
de la Recherche project no. 11-BS04-015-01. The numerical part of this work was performed using the HPC resources from GENCI-CCRT (Grant 2012-t2012056929).
%\bibliographystyle{h-elsevier}
%\bibliography{spires}

\end{document}